\documentclass[aps,twocolumn,showpacs,superscriptaddress,floatfix,prb]{revtex4-2}
\usepackage{amsfonts,hyperref}

\usepackage{graphicx}
\usepackage{amsmath}
\usepackage{amssymb}

\newcommand{\expect}[1]{\mbox{$\langle #1 \rangle$}}

\begin{document}
\title{Superconducting stripes in the hole-doped three-band Hubbard model}

\author{Boris \surname{Ponsioen}} \affiliation{Institute for Theoretical Physics Amsterdam and Delta Institute for Theoretical Physics, University of Amsterdam, Science Park 904, 1098 XH Amsterdam, The Netherlands}
\author{Sangwoo S.~\surname{Chung}} \affiliation{Institute for Theoretical Physics Amsterdam and Delta Institute for Theoretical Physics, University of Amsterdam, Science Park 904, 1098 XH Amsterdam, The Netherlands}
\author{Philippe \surname{Corboz}}  \affiliation{Institute for Theoretical Physics Amsterdam and Delta Institute for Theoretical Physics, University of Amsterdam, Science Park 904, 1098 XH Amsterdam, The Netherlands}

\date{\today}

\begin{abstract}
We study the ground state properties of the hole-doped three-band Hubbard (Emery) model, describing the copper-oxygen planes of the cuprates, using large-scale 2D tensor network calculations. Our simulations reveal a period 4 stripe state with spin and weak charge order over an extended doping range beyond $\delta\sim0.12$ and stripes with larger periods at smaller doping. The period 4 stripe exhibits coexisting $d$-wave superconductivity in the doping range $0.15 \lesssim \delta < 0.25$, while at smaller doping around $\delta \sim 1/8$ we find a strong competition between superconducting and non-superconducting stripes, including also pair density-wave states with alternating sign structure on neighboring stripes, suggesting that the fate of superconductivity around 1/8 doping may be sensitive on the model parameters. 
\end{abstract}

\maketitle

\section{Introduction}

The two-dimensional (2D) Hubbard model is of central importance in the field of strongly correlated systems. In particular since the discovery of high-T$_c$ superconductivity~\cite{bednorz86}, there has been an enormous effort in understanding its phase diagram and its connection to the cuprates~\cite{lee06,keimer15,leblanc15,robinson19,qin22,arovas22}. There is a growing consensus that the basic 2D Hubbard model in the strongly correlated regime does not exhibit superconductivity over an extended doping range, instead, the lowest energy state is a stripe state in which the spin and charge order is modulated with a period 8 at hole density (doping) 1/8~\cite{zheng17}, or more generally stripes with a filling fraction of one hole per unit length ($\rho_l=1$)~\cite{Ido18,tocchio19,qin20}. 

By adding a negative next-nearest neighbor hopping to the model, the periodicity and filling fraction of the stripe decreases~\cite{Ido18,huang18,jiang19,ponsioen19}, leading to period 4  stripes over an extended parameter regime, which is also the typical stripe period observed in the cuprates~\cite{tranquada95,tranquada97,mesaros16}. 
While at doping 1/8 superconductivity was found to be suppressed in the period 4 stripe~\cite{Ido18,ponsioen19}, coexisting $d$-wave superconductivity was observed at larger doping  around $\delta\gtrsim0.14$ ($\rho_l\gtrsim 0.57$)  based on infinite projected entangled-pair states (iPEPS)~\cite{ponsioen19}. In the same study coexisting superconductivity was also predicted in period 5 stripes beyond a similar filling fraction  (corresponding to a doping $\delta\gtrsim0.11$). This is compatible with a recent constrained-path auxiliary-field quantum Monte Carlo (CP-AFQMC) and density matrix renormalization group (DMRG) study, in which coexisting superconductivity in stripes with $\rho_l\ge 0.625$ was revealed~\footnote{At 1/8 doping the ground state was found to be a period 5 instead of a period 4 stripe as predicted by VMC~\cite{Ido18} and iPEPS~\cite{ponsioen19} which is likely due to the smaller $U/t=8$ used in the former, compared to $U/t=10$ in used in the latter two studies. Another VMC study~\cite{marino22} also predicted period 5 stripes at 1/8 doping at $U/t=10$.}. In variational Monte Carlo (VMC) studies, in contrast, no evidence for coexisting superconductivity was found~\cite{Ido18,marino22}. Thus, the relationship between stripe order and superconductivity remains an important topic that calls for further studies.

\begin{figure}[]
  \centering
  \includegraphics[width=\linewidth]{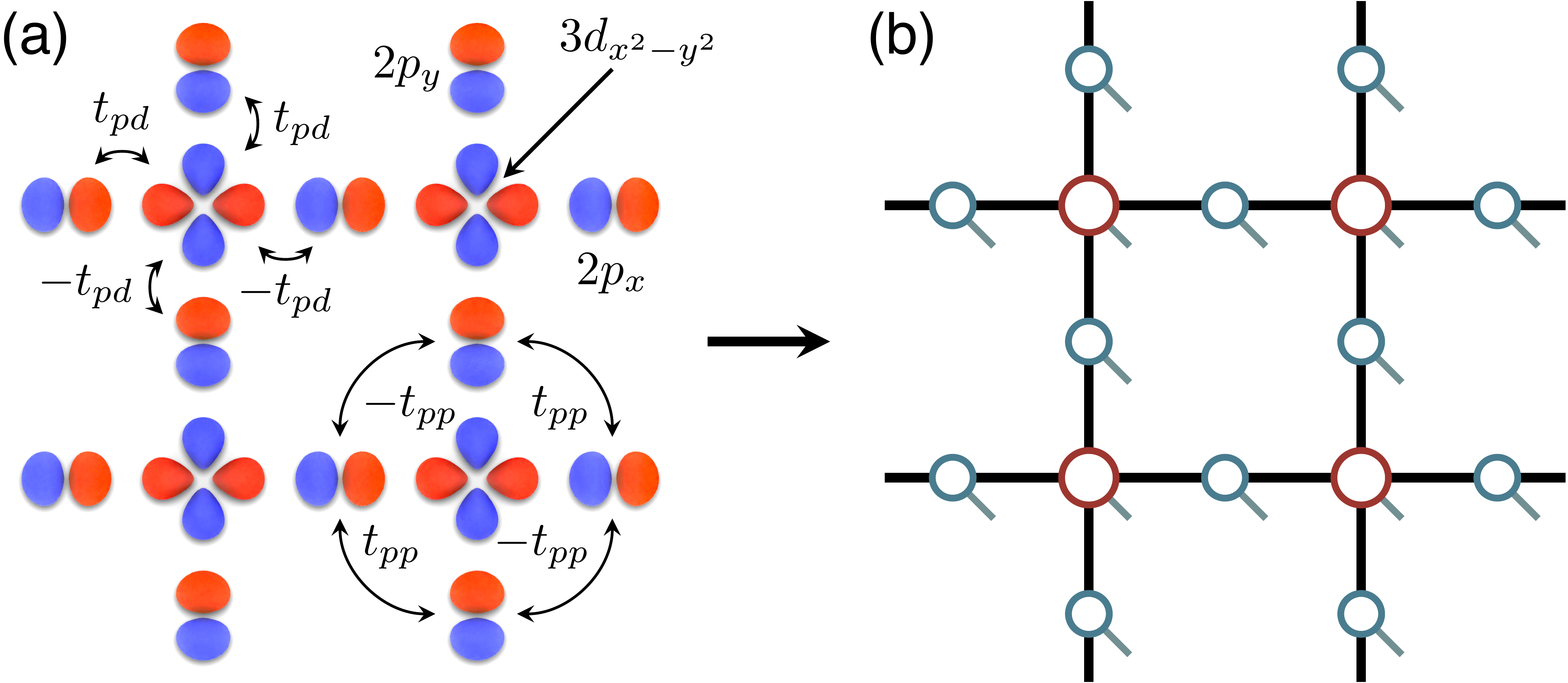} 
  \caption{(a) Copper-oxygen plane with the copper-$3d_{x^2 - y^2}$ orbitals on the vertices and the oxygen $2p_x$ and $2p_y$ orbitals on the links. (b) Corresponding iPEPS ansatz with one tensor (blue and green shapes) per orbital. The open legs correspond to the physical indices of the tensors.}
  \label{fig:ipeps}
\end{figure}

While the single-band Hubbard model captures several features of the cuprate phase diagram, it may be too simplistic to correctly describe the subtle interplay between superconductivity and stripe order. A more realistic description of the cuprates is provided by the three-band Hubbard model (also called Emery-model~\cite{emery87}) which includes the copper $3d_{x^2-y^2}$ orbitals and the oxygen $2p_{x}$ and $2p_{y}$ orbitals as shown in Fig.~\ref{fig:ipeps}. In the hole representation the Hamiltonian can be defined as
\begin{equation}
\label{eq:h}
\begin{split}
& \hat{H} =  \epsilon_d \sum_{i, \sigma} \hat{n}^d_{i\sigma}   + \varepsilon_p \sum_{j, \sigma} \hat{n}^p_{j\sigma}   \\
&  + U_d  \sum_{i}  \hat{n}^d_{i\uparrow}  \hat{n}^d_{i\downarrow}
+ t_{pd} \sum_{\langle i,j\rangle , \sigma} s_{ij} \left(\hat{d}^{\dagger}_{i\sigma} \hat{p}^{}_{j\sigma} + h.c \right) \\
& 
+   t_{pp} \sum_{\langle j,j'\rangle, \sigma}  s_{jj'}  \left(\hat{p}^{\dagger}_{j\sigma} \hat{p}^{}_{j'\sigma} + h.c \right)
  + U_p \sum_{j} \hat{n}^p_{j\uparrow}  \hat{n}^p_{j\downarrow} \,,
\end{split}
\end{equation}
where $\hat{d}^{\dagger}_{i\sigma}$ ($\hat{p}^{\dagger}_{j\sigma}$)  creates a hole with spin $\sigma$ on the copper site $i$ (oxygen site $j$), with $\hat{n}^d_{i\sigma} = \hat{d}^{\dagger}_{i\sigma} \hat{d}_{i\sigma}$ and $\hat{n}^p_{j\sigma} =  \hat{p}^{\dagger}_{j\sigma} \hat{p}_{j\sigma}$ the corresponding number operators. $\epsilon_d$ and $\epsilon_p$ are the site energies and $U_d$ and $U_p$ the on-site interactions on the copper and oxygen orbitals, respectively. The hopping terms with amplitudes $t_{pd}$ and $t_{pp}$ involve sign factors $s$ as indicated in Fig.~\ref{fig:ipeps}. Various parameter sets have been proposed for the different cuprate families based on different techniques~\cite{hybertsen89,mcmahan90,hirayama18,hirayama19,moree22}, some sets including also further-ranged hoppings and interactions. 

The three-band Hubbard model has been subject of many studies in the past decades~\cite{emery87, zaanen88, zaanen89,zhang90,dopf90,scalettar91,guerrero98,yanagisawa01,kivelson04,kent08,arrigoni09,demedici09,yanagisawa09,weber09,hanke10,weber12,weber14,yanagisawa14,kung14,bulut15,go15,white15,thomson15,fratino16,kung16,huang17,ishigaki19,zegrodnik19,dash19,vitali19,ishigaki19,chiciak20,cui20,biborski20,zegrodnik21,mai21,mai21b,chikano21,kowalski21,jiang22}. 
While stripe states on the hole-doped side have previously been found in mean-field studies~\cite{zaanen89}, DMRG ~\cite{white15,jiang22},  CP-AFQMC~\cite{chiciak20}, and determinant QMC~\cite{huang17} (fluctuating stripes), it is still remains a major open question to what extent stripes coexist with superconductivity.

In this paper we study the hole-doped three-band Hubbard model using state-of-the-art 2D tensor network methods designed to simulate the copper-oxygen plane, with a particular focus on the superconducting properties of the ground state. Based on the parameter set from Ref.~\cite{kung16}, we show that the ground state is a period 4 stripe over an extended doping range $\delta \sim 0.12 - 0.25$, while larger stripe periods are obtained at lower doping. Around 1/8 doping our simulations reveal a delicate competition between superconducting and non-superconducting stripes, whereas at larger doping $0.15 \lesssim \delta < 0.25$ we find compelling evidence for coexisting $d$-wave superconductivity. We also discuss the competition between stripes with coexisting in-phase and anti-phase $d$-wave superconducting orders, where the latter (also called pair density-wave state~\cite{berg09,fradkin15,agterberg20}) is found to have a similar or only slightly higher energy than the former one.

\section{Method} 
Our simulations are based on iPEPS - a variational tensor network ansatz to represent 2D wave functions in the thermodynamic limit~\cite{verstraete2004, nishio2004, jordan2008}, which in the past decade has emerged as a powerful tool to study strongly correlated systems, see e.g. Refs.~\cite{Zhao12, Corboz12_su4, corboz14_tJ, corboz14_shastry, niesen17, liao17, chen18, lee18, jahromi18, yamaguchi18, haghshenas19, chung19,kshetrimayum19b, lee20, gauthe20, hasik21, shi22, liu22b}. It can be seen as a natural generalization of matrix product states to higher dimensions, with the main idea to represent quantum many-body states as a trace over a product of tensors.
The ansatz consists of a supercell of tensors that is periodically repeated on the infinite lattice, with typically one tensor per lattice site. For the copper-oxygen plane (corresponding to a Lieb lattice) we use one tensor per orbital, as shown in Fig.~\ref{fig:ipeps}. Each tensor has a physical index carrying the local Hilbert space of a lattice site and four (two) auxiliary indices connecting neighboring tensors on the copper (oxygen) sites, respectively. The number of variational parameters (i.e. the accuracy of the ansatz) is systematically controlled by the bond dimension $D$ of the auxiliary indices. Here we employ the fermionic version of iPEPS~\cite{corboz2010,Corboz09_fmera,kraus2010,Barthel2009} with implemented U(1) spin symmetry~\cite{singh2010,bauer2011} (at half filling we additionally also included the U(1) charge symmetry). Away from half filling, the particle density is controlled by introducing a chemical potential to the Hamiltonian. To obtain quantities at a specific target doping, an interpolation of the data is used.

For an introduction and technical details on iPEPS we refer to Refs.~\cite{corboz2010,phien15,bruognolo21}. For the experts, we note that the contraction of the tensor network is done based on the corner transfer matrix method~\cite{nishino1996, Orus2009} generalized to arbitrary supercell sizes~\cite{Corboz2011,corboz14_tJ} and adapted to the Lieb lattice. The optimization of the iPEPS (i.e. to find the optimal variational parameters) is performed using an imaginary time evolution with a 3-site cluster update~\cite{wang11b,niesen18,ponsioen19}. We found that this approach provides a good trade-off between computational cost and accuracy, enabling us to push the simulations to large bond dimensions up to $D=24$. We also conducted crosschecks at smaller bond dimension based on energy minimization algorithms~\cite{corboz16b,vanderstraeten16,liao19,ponsioen22}. The latter can be pushed to a higher precision for a given bond dimension, but due to a larger computational cost, these calculations are limited to smaller bond dimension.

To identify the low-energy states, simulations using different supercell sizes are performed. The smallest supercell of size $2\times 2$, corresponding to one unit cell of the Lieb lattice, contains 3 tensors. In order to represent stripe states which break the translational symmetry, larger supercell sizes are required. Here we considered sizes up to  $32 \times 4$, enabling to represent stripes with up to a period 8 in the charge order (and period 16 in the spin order).

\section{Results}

\subsection{Benchmark results at half filling} 
\begin{figure}[]
  \centering
  \includegraphics[width=\linewidth]{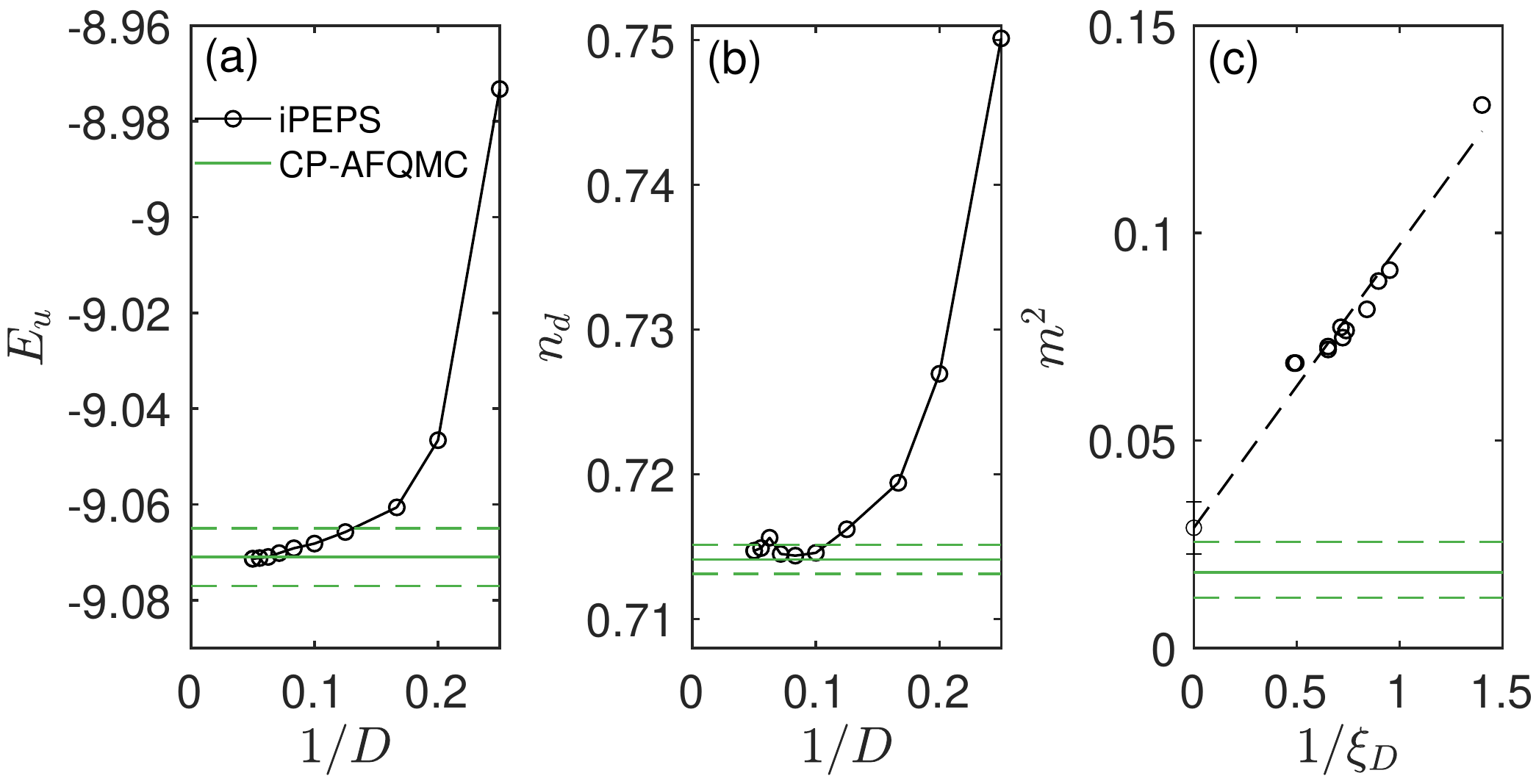}
  \caption{iPEPS data at half filling compared with CP-AFQMC results~\cite{vitali19,chiciak20} for (a) the energy per unit cell, (b) the hole density on the copper site, and (c) the local magnetic moment squared. 
The system sizes used in CP-AFQMC are $12\times 12$ in (a) and (c), and $8\times \infty$ in (b). }
  \label{fig:hf}
\end{figure}

To test the validity of the approach we first present a benchmark comparison at half filling with constrained path auxiliary-field quantum Monte Carlo (CP-AFQMC) from Refs.~\cite{vitali19,chiciak20}, with parameters (in units of eV) $t_{pd}=1.2$, $t_{pp}=0.7$, $U_d=8.4$, $U_p=2$, and charge-transfer energy $\Delta_{pd}=\epsilon_p - \epsilon_d = 4.4$. At half filling, which corresponds to a total hole density per unit cell $n_u=1$, the ground state is insulating and exhibits antiferromagnetic long-range order, with  finite local magnetic moments, $s^z_k = \frac12 \langle \hat n_{k\uparrow} - \hat n_{k\downarrow} \rangle$, residing on the copper sites. 
Figure~\ref{fig:hf}(a)-(b) shows the energy per unit cell $E_u$ and the hole density on the copper site $n_d$ as a function of inverse bond dimension, respectively, where the values at large bond dimension are found to be within the error bars of the CP-AFQMC results. 
The average magnitude of the local magnetic moments, $m$, 
at finite $D$ is higher than the value found in CP-AFQMC on the $12\times12$ lattice~\cite{vitali19} (Fig.~\ref{fig:hf}(c)). However, by extrapolating the iPEPS to the infinite bond dimension limit, we find a value for $m^2$ which seems compatible with the CP-AFQMC result. Here we employed an extrapolation based on the effective correlation length $\xi_D$ extracted from the iPEPS at each value of $D$, as previously done for the 2D Heisenberg model in Refs.~\cite{corboz18,rader18,hasik21}. Thus, we conclude that iPEPS and CP-AFQMC provide compatible results at half filling.

\subsection{Ground state at finite hole doping}
\begin{figure}[]
  \centering
  \includegraphics[width=\linewidth]{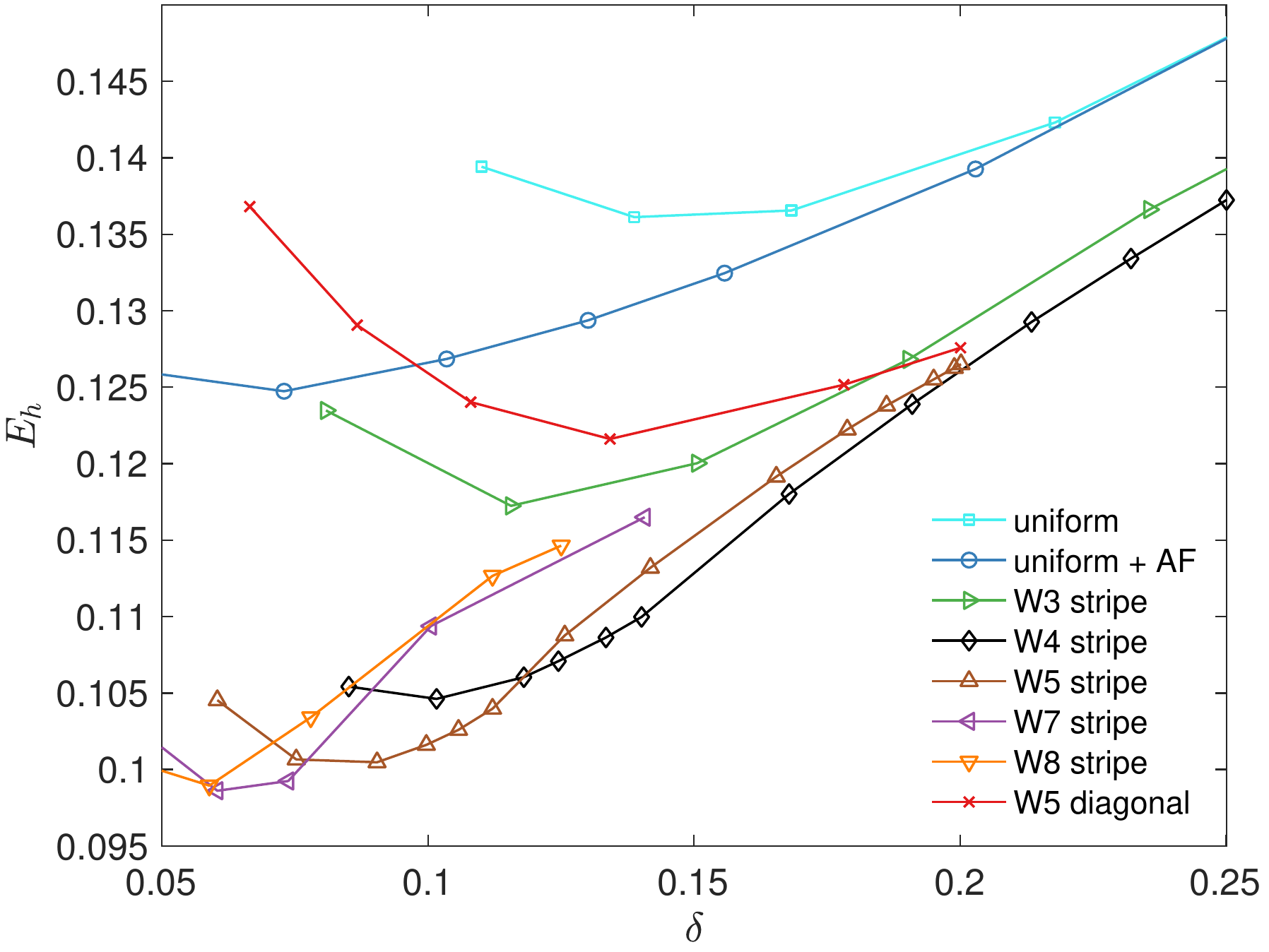}
  \caption{Energy per hole as a function of hole doping $\delta$ of various competing low-energy states obtained with different supercells ($D=18$), including stripe states with periods between 3 and 8 (W3-W8), a diagonal stripe state with period 5, and uniform states (i.e. without charge order) with and without coexisting antiferromagnetic (AF) order, respectively. In the doping range $\delta \sim 0.1\ldots 0.2$ the period 5 and period 4 stripes exhibit the lowest variational energy. }
  \label{fig:Eh}
\end{figure}

We next consider the hole-doped case, using the parameter set from Ref.~\cite{kung16}, $t_{pd}=1.13$, $t_{pp}=0.49$, $U_d=8.5$, $U_p=4.1$, and $\Delta_{pd}=\epsilon_p - \epsilon_d = 3.24$ (in units of eV), which is considerably more challenging than the half-filled case. 
 As in the one-band Hubbard model~\cite{zheng17,ponsioen19} we find a competition between different stripe and uniform states which are obtained using iPEPS with different supercell sizes. Figure~\ref{fig:Eh}  shows the energy per hole, $E_h(\delta) = [E_u(\delta) - E_u(\delta=0)]/\delta$, of the main competing states for $D=18$, with $\delta$ the hole doping (i.e. hole density per unit cell with respect to half filling, $\delta = n_u-1$). We find that in the doping range of $\delta\sim 0.12 \ldots 0.25$ the period 4 (W4) has the lowest variational energy. The stripe period of the lowest energy state increases with decreasing doping, as in the one-band case, requiring larger and larger cell sizes. Here we  focus  on results beyond 10\% doping. Uniform states (i.e. without stripe order) as well as diagonal stripes are found to be  higher in energy than the vertical stripe states.
Qualitatively similar results are also found for other bond dimensions and for simulations based on the energy minimization algorithm, see supplemental material~\cite{SM}.

\begin{figure}[t]
  \centering
    \includegraphics[width=\linewidth]{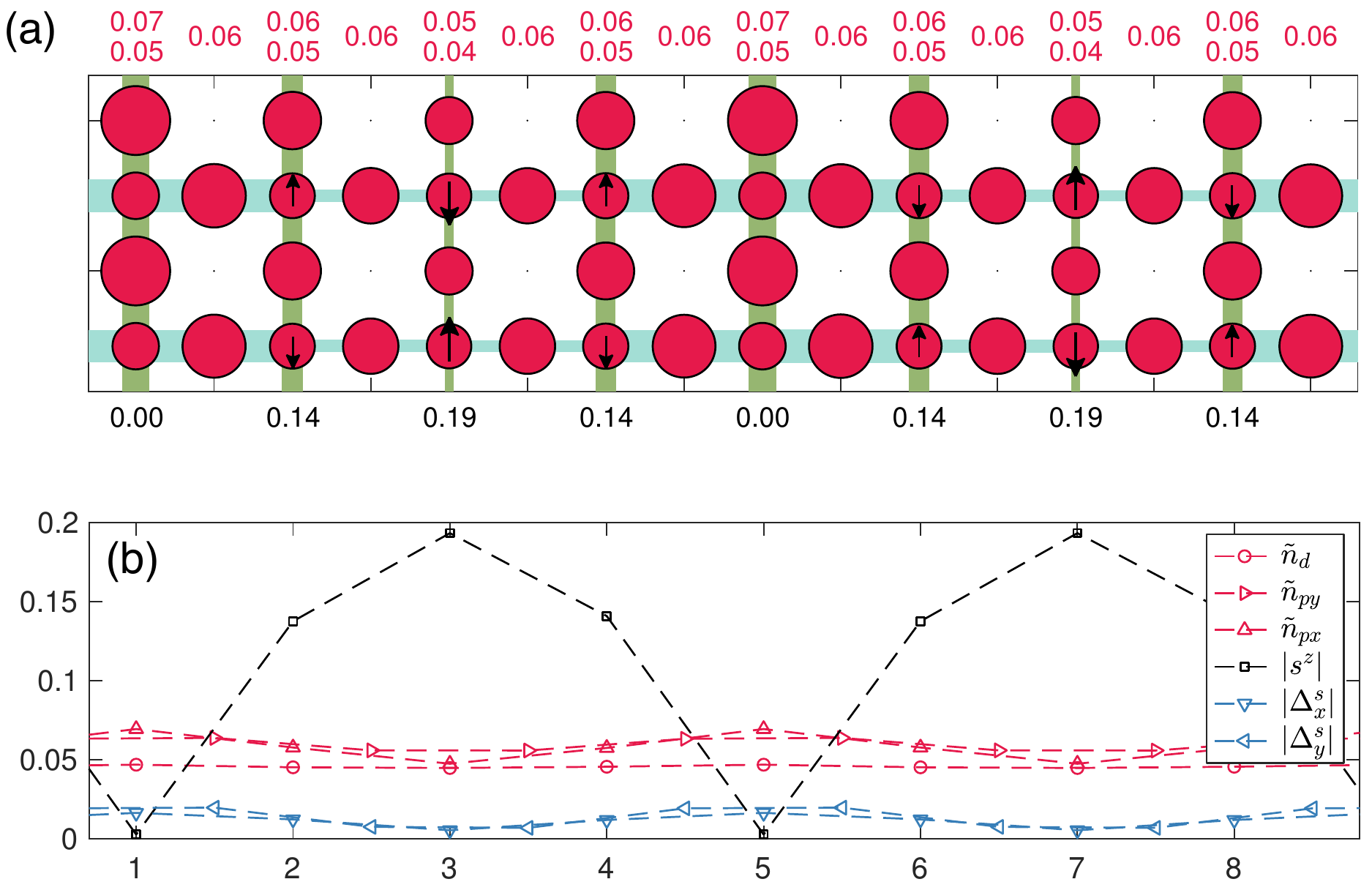} 
  \caption{(a) Example of a stripe with period 4 (W4) in the charge order obtained with iPEPS using a $16\times4$ supercell with 48 tensors at doping $\delta\sim0.16$ ($D=20$). 
  The sizes of the black arrows and the red discs are proportional to the local magnetic moment (formed on the copper sites) and the local hole density, respectively, where for the latter the values at half filling have been subtracted. The magnitudes averaged over a column of equivalent sites are indicated in the bottom and top rows. 
  The width of a bond is proportional to the  singlet pairing amplitude between neighboring copper sites, with different signs in x- and y- direction represented by the two different colors.   (b) A plot of the corresponding magnitudes of the values shown in (a), where the red symbols show the hole densities with the values at half filling subtracted, $\tilde n_\alpha = n_\alpha(\delta) - n_\alpha(n_u=1)$, with $\alpha \in \{d, p_x, p_y\}$ for the  copper $3d_{x^2-y^2}$, oxygen $p_x$ and $p_y$ orbital sites, respectively. The black squares display the magnitude of the local spin moment, $|s^z|$, and the blue symbols show the magnitude of the local singlet pairing amplitude in x- and y- direction, $|\Delta^s_x |$ and $|\Delta^s_y|$.
 }
  \label{fig:period4stripe}
\end{figure}

An example of a period 4 stripe around doping $\delta\sim 0.16$  is shown in Fig.~\ref{fig:period4stripe}, displaying the local hole density and local magnetic moment  on each site, as well as the singlet pairing amplitude between neighboring copper sites, $\Delta^s_{ii'}=\langle \hat c_{i\uparrow} \hat c_{i'\downarrow} - \hat c_{i\downarrow} \hat c_{i'\uparrow}  \rangle / \sqrt{2}$. The charge order predominantly develops on the p-orbitals, with a slightly stronger modulation on the $p_y$ orbitals than on the $p_x$ orbitals (for a vertical stripe). The magnetic moments only form on the copper sites, where the spin order exhibits the characteristic $\pi$-phase shift across the column of sites with maximal hole density. Due to the phase shift the period of the spin order is doubled for stripes with an even charge period. These observations are in agreement with previous findings~\cite{white15,chiciak20}. 

\begin{figure}[t]
  \centering
  \includegraphics[width=\linewidth]{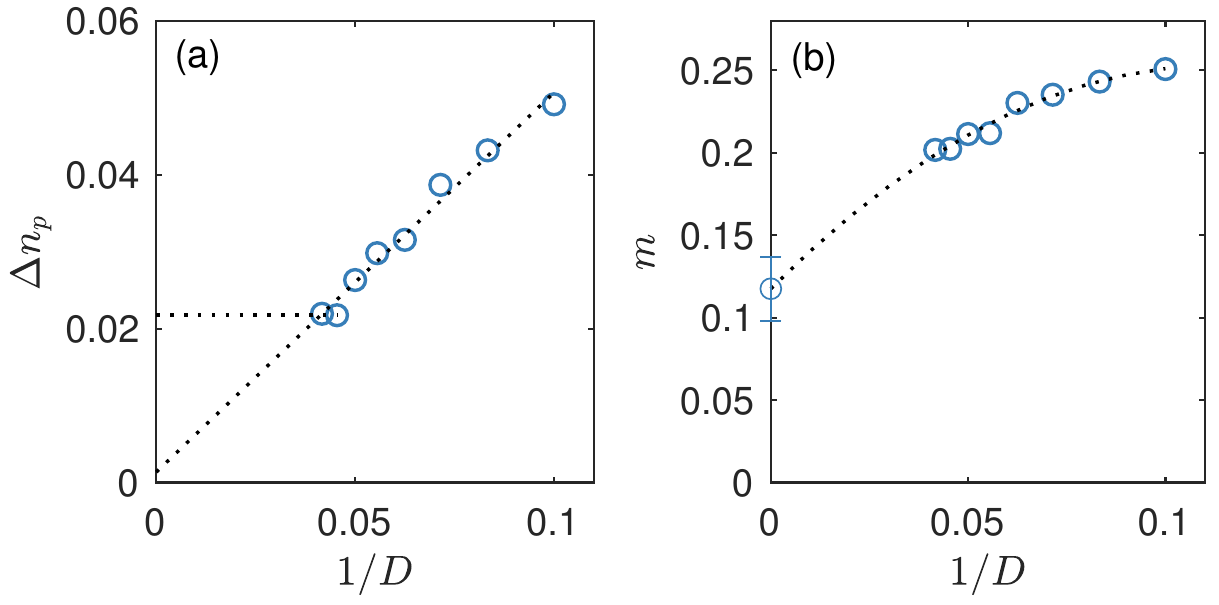}
  \caption{(a) Amplitude of the charge order, $\Delta n_p = \max_j {\sum_\sigma \expect{\hat{n}^p_{j\sigma}}} -  \min_j {\sum_\sigma \expect{\hat{n}^p_{j\sigma}}}$, and (b) maximum of the local magnetic moment of the period 4 stripe at doping $\delta=1/8$ as a function of inverse bond dimension.
  }
  \label{fig:props}
\end{figure}

In Fig.~\ref{fig:props}(a) and (b) the strength of the charge and spin orders are shown as a function of inverse bond dimension. The charge order is  weak already at finite $D$, and tends to a very small or even vanishing value in the infinite $D$ limit. On the other hand, the charge densities may also saturate at finite $D$, similarly as in Fig.~\ref{fig:hf}(b), at a value in between 0 and 0.02. The maximum of the local magnetic moment is also suppressed with increasing $D$, but it clearly remains finite in the infinite bond dimension limit.

\subsection{Superconductivity in the period 4 stripe}
\begin{figure}[]
  \centering
  \includegraphics[width=\linewidth]{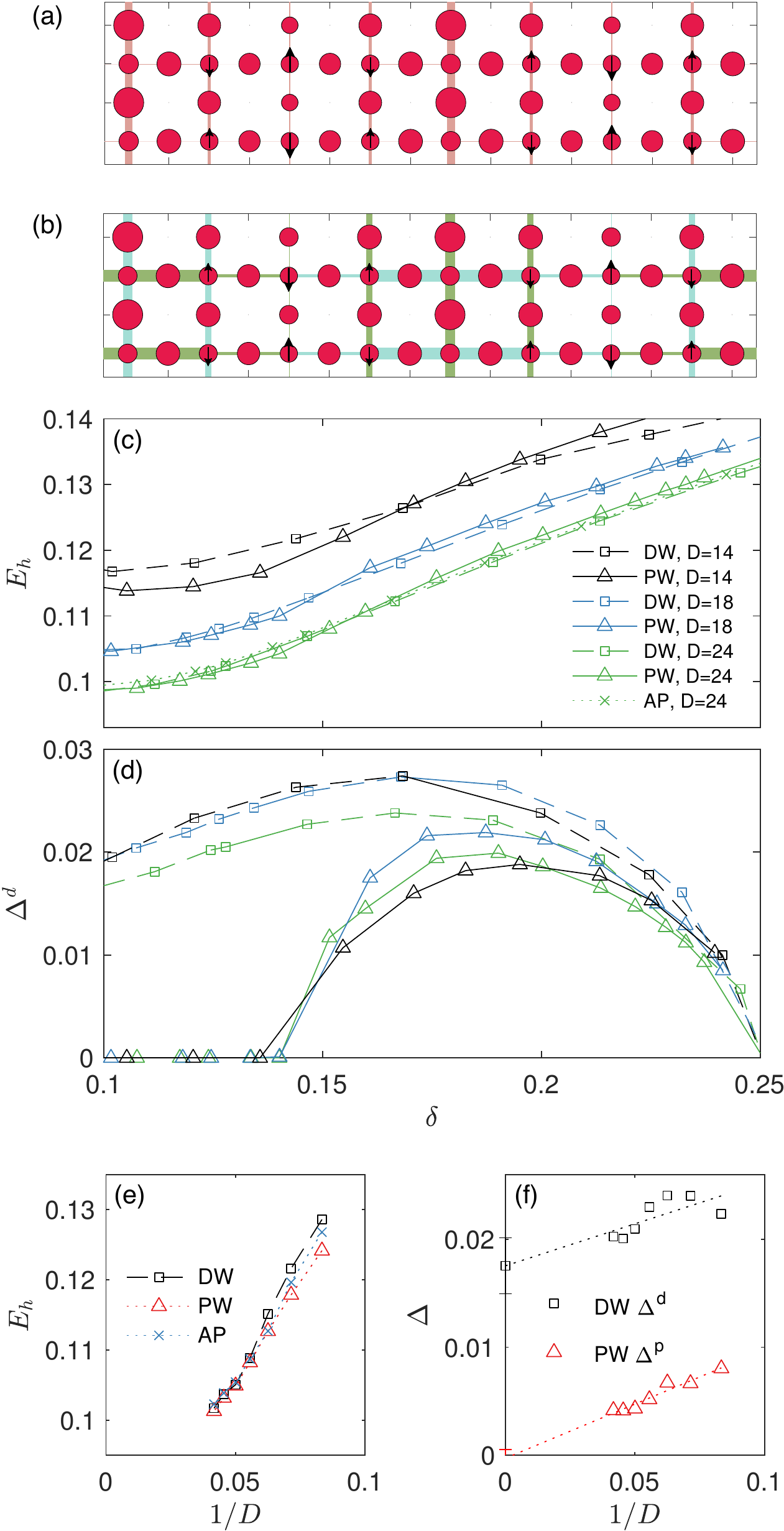}
  \caption{
  (a) Period 4 stripe with suppressed $d$-wave pairing but with (short-range) triplet pairing ($p$-wave) along the vertical direction. 
  (b) Competing period 4 stripe with anti-phase superconducting order (pair density-wave state). 
  (c)~Energies of the period 4 stripes with $d$-wave (DW), short-range p-wave (PW), and anti-phase $d$-wave (AP) order as a function of doping, revealing a strong competition between the states.
  (d)~Average $d$-wave pairing strength of the DW and PW states as a function of doping. The former exhibits coexisting $d$-wave superconductivity over the entire doping range (except at 0.25 where the stripe is fully filled). In the latter $d$-wave superconductivity is entirely suppressed below $\delta \sim 0.15$ but finite at larger doping. (e) Energies of the competing period 4 stripes as a function of inverse $D$ at 1/8 doping, showing that the energies are very close at large $D$. (f) Magnitude of the $d$-wave pairing of the DW state and triplet pairing of the PW state as a function of inverse $D$ at 1/8 doping. The former is finite in the infinite $D$ limit whereas the latter vanishes. 
  }
  \label{fig:sc}
\end{figure}

We next turn our focus on the pairing properties of the stripe states, which has not been addressed yet with AFQMC for the 3-band model, and previous DMRG studies have been limited to small-width cylinders only~\cite{white15,jiang22}.
The simulations reveal three different competing period 4 stripes which are very close in energy (see Figs.~\ref{fig:sc}(c),(e)) but with different pairing properties~\footnote{The different states have been obtained by starting from different (random) initial states, i.e. each state corresponds to a local minimum obtained through the optimization of the tensors.}. We first discuss two of the states and then comment on the third one further below. The first state exhibits coexisting $d$-wave superconductivity (like in Fig.~\ref{fig:period4stripe}) over the entire doping range, except for the fully filled stripe at $\delta = 0.25$. We measure its strength by the pairing amplitude $\Delta^d$, defined as the site-averaged $\Delta^s_{ii'}$ with different sign factors in x and y direction, see Fig.~\ref{fig:sc}(d).
In the second state $d$-wave superconductivity is entirely suppressed below $\delta \sim 0.15$, but instead, it exhibits a finite triplet pairing ($p$-wave) between neighboring copper sites, $\Delta^t_{ii'}=\langle \hat c_{i\uparrow} \hat c_{i'\downarrow} + \hat c_{i\downarrow} \hat c_{i'\uparrow}  \rangle / \sqrt{2}$ along the stripe at finite $D$, as displayed in Fig.~\ref{fig:sc}(a). 
However, the triplet pairing strength $\Delta^p$, given by the site-averaged $\Delta^t_{ii'}$ (with equal signs in x and y direction), vanishes in the infinite $D$ limit (Fig.~\ref{fig:sc}(f)), i.e. the stripe is non-superconducting. 
At larger doping $0.15 < \delta < 0.25$ both states have coexisting $d$-wave superconductivity as shown in Fig.~\ref{fig:sc}(d), albeit with different magnitudes of the pairing strength.

Thus, while we find clear evidence of superconducting stripes beyond $\delta\sim 0.15$, corresponding to a stripe filling fraction of $\rho_l = W\cdot \delta  \sim 0.6$, the situation at lower doping around $1/8$ doping is less clear, since the two competing states are nearly degenerate in the large bond dimension limit (Fig.~\ref{fig:sc}(e)). 
A similar competition between superconducting and non-superconducting stripes can also be found for the period 5 stripe, where the onset of $d$-wave superconductivity appears at a similar filling fraction $\rho_l\sim 0.6$, corresponding to a lower doping of $\delta\sim 0.12$, see supplemental material~\cite{SM}. 

A strong competition between short-range $p$-wave pairing and a SC $d$-wave stripe was also predicted in the single band $t$-$t'$ Hubbard model~\cite{ponsioen19} for $U/t=10$, where a transition between the two states was found to occur around $\delta\sim0.14$ for the period 4 stripe, and $\delta\sim0.115$ for the period 5 stripe (corresponding to $\rho_l\sim0.57$) which are close to above values. 
Evidence for coexisting superconductivity in stripes with a filling fraction between $\rho_l = 0.6\ldots 0.8$ was also found in Ref.~\cite{xu23} at $U/t=8$.

Finally, we discuss the third competing state, which exhibits coexisting $d$-wave superconductivity  but with alternating signs  on neighboring stripes as shown in Fig.~\ref{fig:sc}(b), also called pair density-wave (PDW) state~\cite{berg09,fradkin15,agterberg20}. This state has been proposed as a possible reason for the absence of 3D superconductivity above $T=4$ K in La$_{2-\delta}$Ba$_{\delta}$CuO$_4$ around $\delta=1/8$~\cite{li07,huecker11}, because the anti-phase pattern leads to a suppression of the interlayer Josephson coupling between the copper-oxygen planes~\cite{berg07}. We find that this state has a similar energy or only slightly higher energy than the other two states, i.e. the energy cost to induce the phase shift in the superconducting order is very small. This is compatible with previous results for the single-band case ($t$-$J$ model)~\cite{himeda02,corboz14_tJ}.  We note that PDW states have recently also been found in DMRG calculations of a three-band Hubbard model on a two-leg square cylinder~\cite{jiang22}.

\section{Conclusions}
Using large-scale  iPEPS  simulations of the hole-doped three-band Hubbard model we have found period 4 stripes over an extended doping range $\delta \sim 0.12 - 0.25$, and longer-period stripes at lower doping. 
The stripes exhibit spin order including the typical $\pi$-phase shift and only weak charge order. Unlike in the basic one band Hubbard model with $t'=0$, the stripes exhibit coexisting $d$-wave superconductivity over an extended doping range beyond $\delta \sim 0.15$. At smaller doping our simulations revealed a strong competition between superconducting and non-superconducting stripes, reminiscent of the competition which was previously observed in the $t-t'$ Hubbard model. The non-superconducting stripes have a lower energy at small $D$, but the energies become nearly degenerate at large $D$, such that the fate of  $d$-wave superconductivity around 1/8 doping may be sensitive to the parameter set used. Finally, we also found competing PDW states with a similar or only slightly higher energy at large $D$. Exploring the dependence of this competition on the various model parameters will be an interesting direction for future research, in order to get further insights into the intriguing interplay between stripe order and superconductivity.

\begin{acknowledgments}
We thank E. Vitali for sharing CP-AFQMC data and for useful discussions. This project has received funding from the European Research Council (ERC) under the European Union's Horizon 2020 research and innovation programme (Grant Agreement No. 677061 and No. 101001604). This work was carried out on the Dutch national e-infrastructure with the support of SURF Cooperative, and is part of the D-ITP consortium, a program of the Netherlands Organization for Scientific Research (NWO) that is funded by the Dutch Ministry of Education, Culture and Science (OCW).
\end{acknowledgments}

\bibliographystyle{apsrev4-2}
\bibliography{../../bib/refs.bib,mhub_note}

\end{document}


\title{Superconducting stripes in the hole-doped three-band Hubbard model: \emph{Supplemental material}}

\author{Boris \surname{Ponsioen}} \affiliation{Institute for Theoretical Physics Amsterdam and Delta Institute for Theoretical Physics, University of Amsterdam, Science Park 904, 1098 XH Amsterdam, The Netherlands}
\author{Sangwoo S.~\surname{Chung}} \affiliation{Institute for Theoretical Physics Amsterdam and Delta Institute for Theoretical Physics, University of Amsterdam, Science Park 904, 1098 XH Amsterdam, The Netherlands}
\author{Philippe \surname{Corboz}}  \affiliation{Institute for Theoretical Physics Amsterdam and Delta Institute for Theoretical Physics, University of Amsterdam, Science Park 904, 1098 XH Amsterdam, The Netherlands}

\date{\today}

\maketitle

In this supplemental material we provide additional results for the energies of the states at different bond dimensions and regarding the competition between superconducting and non-superconducting period 5 stripes.

\begin{figure}[t]
  \centering
  \includegraphics[width=\linewidth]{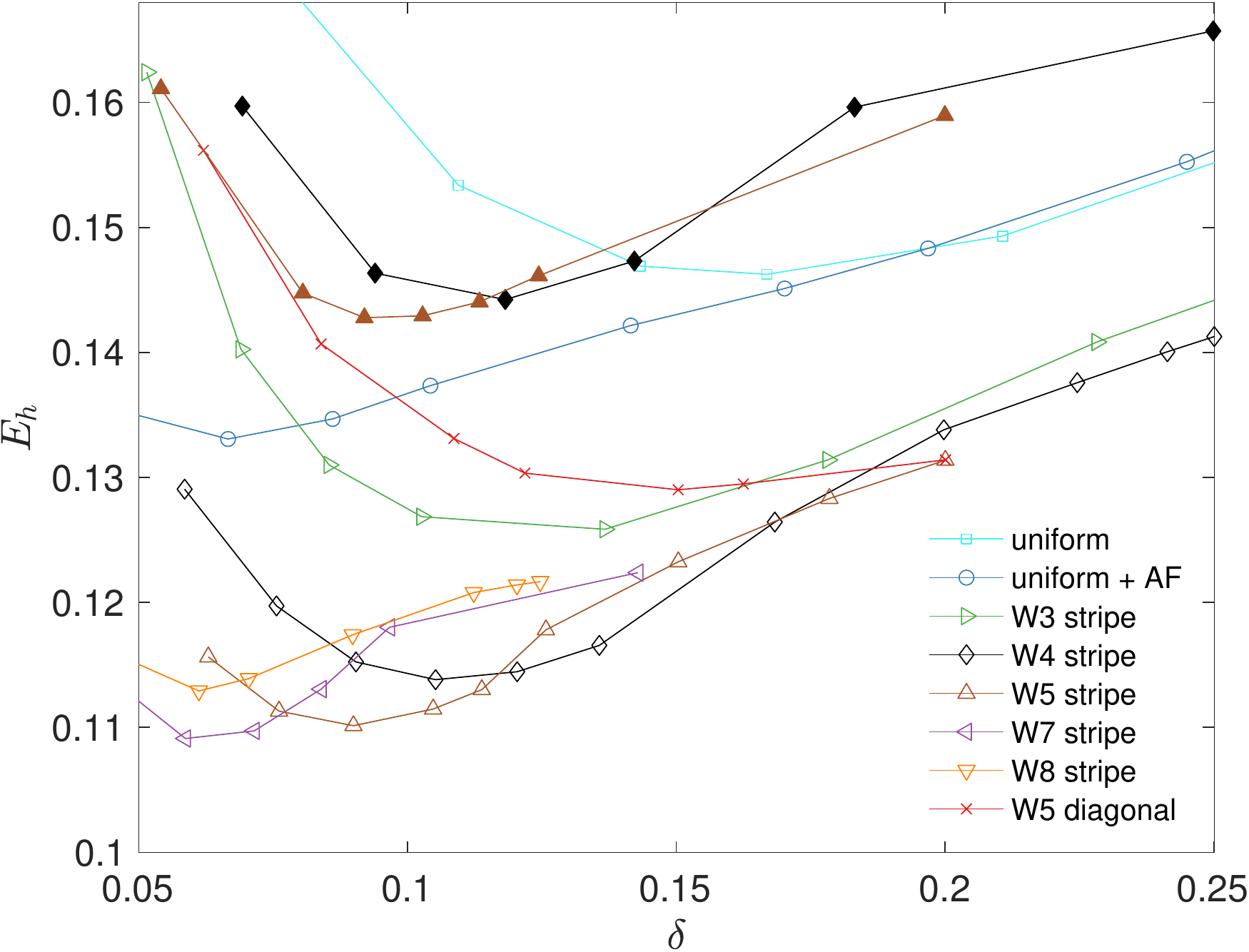}
  \caption{
  Energy per hole as a function of hole doping $\delta$ of various competing low-energy states obtained with iPEPS with different supercells, similar as in Fig.~3 in the main text, but here for $D=14$ (open symbols) obtained with imaginary time evolution. Additionally, $D=8$ data obtained with automatic differentiation is shown (filled symbols). Also here we find that in the doping range $\delta \sim 0.1\ldots 0.2$ the period 4 and period 5 stripes exhibit the lowest  energy. }
  \label{fig:Eh}
\end{figure}

Figure~\ref{fig:Eh} shows the energy per hole as a function of the competing states for a bond dimension $D=14$, and additional data for $D=8$ which was obtained by energy minimization based on variational optimization~\cite{corboz16b} and further refined using automatic differentiation~\cite{liao19}, implemented for fermionic iPEPS with U(1) symmetry~\cite{ponsioen22}. Qualitatively the results are similar to the ones in Fig. 3 of the main text at larger bond dimension $D=18$, i.e. the period 4 stripe is the lowest energy state over an extended doping range beyond $\delta \sim 0.12$, and stripes with longer periods are found at lower doping. One quantitative difference is that here the period 5 stripe is slightly lower around $\delta\sim 0.2$, which is not the case at higher bond dimensions.

\begin{figure}[h]
  \centering
  \includegraphics[width=\linewidth]{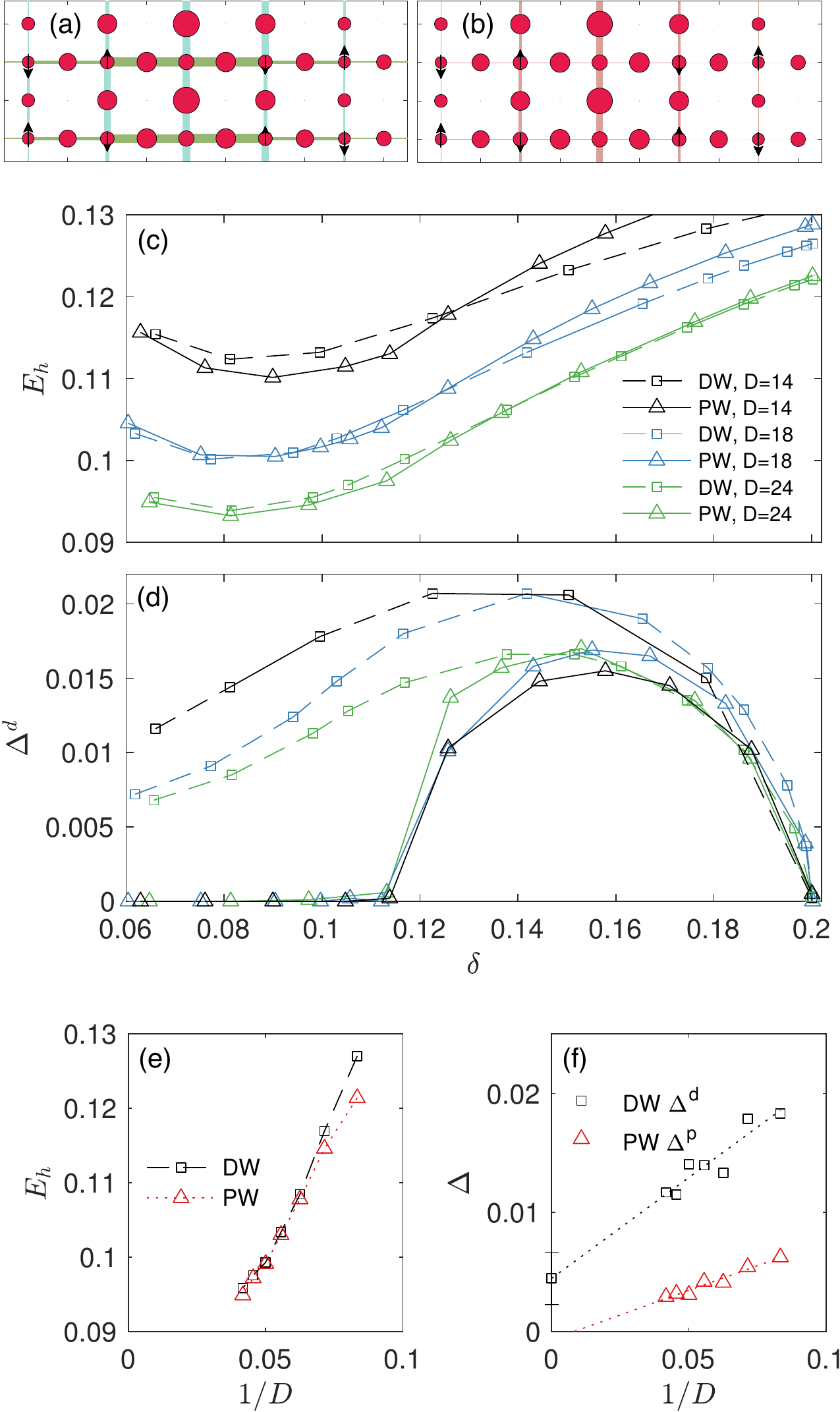}
  \caption{Results for the two competing period 5 stripes with different pairing properties: one with coexisting $d$-wave pairing over the entire doping region shown in (a) and the other one with vanishing $d$-wave pairing but short-ranged triplet pairing ($p$-wave) at doping below $\delta\sim 0.12$, shown in (b). 
  (c)~Energies of the two states which are very close at high bond dimension. (d) Average $d$-wave pairing which is finite in both states beyond $\delta\sim 0.12$.  (e) Energies of the two states at $\delta=1/10$ as a function of inverse $D$, showing that the energies are very close at large $D$. (f) Magnitude of the $d$-wave ($p$-wave) pairing of the DW (PW) state at $\delta=1/10$ as a function of inverse $D$ which is finite (vanishing) in the infinite $D$ limit. }
  \label{fig:sc}
\end{figure}

In Fig.~\ref{fig:sc} we present results for the competition between superconducting and non-superconducting period 5 stripes, showing a very similar behavior as in the case of the period 4 stripe in Fig. 6 in the main text.
 The state shown in Fig.~\ref{fig:sc}(a) exhibits coexisting $d$-wave superconductivity over the entire doping range shown in Fig.~\ref{fig:sc}(d), except at doping 0.2 when the stripe is fully filled ($\rho_l=1$). In the second state shown in Fig.~\ref{fig:sc}(b) $d$-wave superconductivity is entirely suppressed below $\delta \sim 0.12$, instead, the state exhibits a finite triplet pairing on the vertical bonds, but without long-range order in the infinite $D$ limit, as shown in Fig.~\ref{fig:sc}(f). Beyond $\delta \sim 0.12$ also this state develops coexisting $d$-wave superconductivity. The onset of $d$-wave superconductivity occurs at a similar filling fraction $\rho_l = W\delta = 0.6$ as in the period 4 case (corresponding to a lower doping due to the larger period), suggesting that the onset depends on the filling fraction. We expect a similar behavior also for the stripes with larger periods. The two competing states are energetically very close, as shown in Fig.~\ref{fig:sc}(c) and (e). Similarly as in the period 4 case we also expect that stripes with anti-phase superconducting order (pair density-wave states) are very close in energy (these would require twice the cell size of the in-phase superconducting stripe). 

\bibliographystyle{apsrev4-2}
\bibliography{../bib/refs.bib,biblio}